\documentclass[12pt]{article}
\usepackage{amsmath,amssymb}
\def\R{\hbox{{\rm I}\kern-0.2em{\rm R}\kern0.2em}}%definition of reals
\def\R{\hbox{{\rm I}\kern-0.2em{\rm R}\kern0.2em}}%mathematical R for reals
\def\s{\hbox{{\rm \subset}\kern-0.2em{\rm +}\kern0.2em}}%for semidirect product
\def\D{\hbox{{\rm I}\kern-0.2em{\rm D}\kern0.2em}}

\def\be{\begin{equation}}
\def\ee{\end{equation}}

\def\({\left(}
\def\){\right)}
\def\[{\left[}
\def\]{\right]}
\def\bc{\begin{center}}
\def\ec{\end{center}}

\begin{document}

{\large \bf Approximate Noether Symmetries of the Geodesic Equations
for the Charged-Kerr Spacetime and Rescaling of Energy}

\textit{IBRAR HUSSAIN}$^\dag$\footnote{Correspondence should be
addressed to ihussain@camp.edu.pk}, \textit{F. M. MAHOMED}$^\ddag$
and \textit{ASGHAR QADIR}$^{\dag}$

$^{\dag}$Centre for Advanced Mathematics and Physics\\
National University of Sciences and Technology\\
Campus of the College of Electrical and Mechanical Engineering\\
Peshawar Road, Rawalpindi, Pakistan

E-mail: ihussain@camp.edu.pk, aqadirmath@yahoo.com

$^{\ddag}$Centre for Differential Equations, Continuum Mechanics and Applications\\
School of Computational and Applied Mathematics\\
University of the Witwatersrand\\
Wits 2050, South Africa

E-mail: Fazal.Mahomed@wits.ac.za

{\bf Abstract}. Using approximate symmetry methods for differential
equations we have investigated the exact and approximate symmetries
of a Lagrangian for the geodesic equations in the Kerr spacetime.
Taking Minkowski spacetime as the exact case, it is shown that the
symmetry algebra of the Lagrangian is 17 dimensional. This algebra
is related to the 15 dimensional Lie algebra of conformal isometries
of Minkowski spacetime. First introducing spin angular momentum per
unit mass as a small parameter we consider first-order approximate
symmetries of the Kerr metric as a first perturbation of the
Schwarzschild metric. We then consider the second-order approximate
symmetries of the Kerr metric as a second perturbation of the
Minkowski metric. The approximate symmetries are recovered for these
spacetimes and there are no non-trivial approximate symmetries. A
rescaling of the arc length parameter for consistency of the trivial
second-order approximate symmetries of the geodesic equations
indicates that the energy in the charged-Kerr metric has to be
rescaled and the rescaling factor is $r$-dependent. This rescaling
factor is compared with that for the Reissner-Nordstr\"{o}m metric.

\textit{Key words}: Kerr; charged-Kerr spacetimes; perturbed
Lagrangian; first-; second-order approximate symmetries; energy

\section{Introduction}

In general a spacetime may not be stationary (and especially may not
be static) and hence local (global) energy conservation may be lost.
Due to this fact there is a long standing problem of the definition
of energy (or mass) in general relativity \cite{MTW}. If the
spacetime is static there is a timelike \textit{isometry} or
\textit{Killing Vector} (KV). Further energy conservation in a
spacetime is guaranteed in the frame using a timelike KV to define
the time direction. However, in the absence of a timelike KV the
energy of a test particle is not defined and hence the energy in the
gravitational field is not well defined. (Of course, one could use
the quasilocal energy defined for a Lagrangian for a {\it field
theory} using an ADM foliation see references \cite{LMMM1, LMM2}).

if there does not exist a timelike KV, energy is not conserved.
Since gravitational wave spacetimes are non-static vacuum solutions
of the Einstein Field Equations (EFEs), for which a timelike KV does
not exist, the  problem of defining the energy content of
gravitational waves is particularly severe. Different people have
tried different \textit{approximate symmetry} approaches
\cite{AK,JM} to define the energy content of gravitational waves but
there is no clear solution to the problem. We use approximate
symmetry methods for differential equations (DEs) \cite{Ib} with the
hope of finding approximate timelike KVs to look at the solution of
the problem. It is obvious that we need to learn how to physically
interpret the results that will emerge from the approximate symmetry
calculations. For this purpose first the approximate symmetries of
the Schwarzschild metric were investigated \cite{KMQ}; next we
studied the Reissner-Nordstr\"{o}m (RN) metric \cite{IMQ}; and here
we consider the Kerr metric. We compare our results for the energy
with those of Komar \cite{AK1} and discuss the difference. In a
subsequent paper we plan to investigate the approximate symmetries
of time-varying spacetimes and hence try to identify what this
approach would give as the energy content of gravitational waves.

The 10 generators of the Poincar$\grave{e}$ isometry algebra
$so(1,3)\oplus _{s}\mathbf{\R}^{4}$, (where $\oplus_{s}$ denotes
semi-direct sum) for the Minkowski spacetime (which is maximally
symmetric) \cite{ESEFEs,AQs} gives conservation laws for energy,
linear momentum and spin angular momentum. Going from Minkowski to
non flat spacetimes like Schwarzschild, RN and Kerr spacetimes some
of the conservation laws are lost because of the gravitational
field. Using Lie symmetry methods \cite{Lie}, first-order
approximate symmetries of the system of the geodesic equations for
the Schwarzschild metric were discussed in \cite{KMQ} and
second-order approximate symmetries of the system of the geodesic
equations for the RN metric were given in \cite{IMQ}. For the
first-order and also for the second-order approximate symmetries,
the lost conservation laws of spin angular momentum and linear
momentum are recovered as trivial approximate conservation laws. In
the case of second-order approximate symmetries of the RN spacetime
one finds that it is necessary to rescale the energy of test
particles.

In this paper we start by using symmetries of the {\it Lagrangian},
rather than those of the geodesic equations. In particular we explore
first and second-order approximate symmetries of a Lagrangian of
the Kerr spacetime. First, we consider the Kerr metric as a first
perturbation of the Schwarzschild metric with spin as a small parameter,
$\epsilon$. The isometry algebra for the Schwarzschild spacetime
\cite{ESEFEs} is $so(3)\oplus$ \R while the symmetry algebra for the
Lagrangian is $so(3)\oplus$\R$\oplus d_{1}$ (where $d_1$ is the Lie
algebra generated by ${\partial}/{\partial s}$). Retaining terms of
first order in $\epsilon$ and neglecting its higher powers we show that
there is no ``non-trivial" (in the technical sense explained in the next
section) first-order approximate symmetry for the Lagrangian of this
perturbed Schwarzschild metric. We only recover the two symmetry
generators of angular momentum as ``trivial" first-order approximate
symmetry generators which were lost in going from Schwarzschild to the
Kerr spacetime. We then consider the Kerr metric as a second perturbation
of the Minkowski metric. Taking Minkowski spacetime as an exact case we
obtain a seventeen dimensional Lie algebra, which contains the ten
dimensional isometry algebra (Poincar\'{e} algebra). The significance of
the remaining seven symmetry generators will be discussed in section 3.
Regarding mass as a small parameter, $\epsilon$, for the approximate
Schwarzschild metric as a first perturbation of the Minkowski spacetime,
we recover all the lost symmetries as ``trivial" first-order approximate
symmetries. The isometry algebra of the unperturbed Kerr spacetime is two
dimensional \cite{ESEFEs} and the symmetry algebra of the Lagrangian for
this spacetime is three dimensional, i.e. the two KVs
${\partial}/{\partial}t$, ${\partial}/{\partial}{\phi}$ and the
translation in the geodetic parameter ${\partial}/{\partial}s$. Now
introducing the spin as a small parameter, $\epsilon$ and retaining
terms of order $\epsilon^{2}$ in the approximate Kerr spacetime as
second perturbation of the Minkowski spacetime we recover all the
lost symmetries of the Lagrangian as ``trivial" second-order
approximate symmetries.

A problem arises in the search for a scaling factor for the energy
of test particles in the Kerr metric. Whereas, in the RN-case the
energy rescaling was by $(1-Q^2/2Gm^2)$, there is a simple
multiplicative factor for the Kerr metric. In the absence of the
constant (unity in this case), it is not clear what significance to
attach to the rescaling. So as to relate that factor to the factor
arising in the RN-case, we investigate second-order approximate
symmetries of the geodesic equations for the charged-Kerr spacetime.
For this purpose we take mass, charge and angular momentum per unit
mass as small parameters, of order $\epsilon$, and only retain the
second power, neglecting its higher powers. More specifically, in
the set of determining equations for second-order approximate
geodesic equations, the coefficient of ${\partial}/{\partial s}$ (in
the point transformation generator given in section 2) collects a
rescaling factor (given in section 4). Since $s$ is the proper time
and energy conservation is related to time translation, the energy
of a test particle in the charged-Kerr spacetime rescales. This
scaling factor consists of two terms, one due to charge and the
other due to the spin of the gravitating source. We then compare
this scaling factor with that of the RN spacetime. We also give a
comparison of the scaling factor obtained here with the already
existing expressions in the literature \cite{CF, DC, AJ} for the
mass (energy) of the charged-Kerr spacetime.

The plan of the paper is as follows. In the next section we briefly
review the definitions of symmetries and approximate symmetries of a
Lagrangian. In section 3, approximate symmetries of the Lagrangian
for the Kerr spacetime are considered. In section 4 we briefly
discuss second-order approximate symmetries of the geodesic
equations for the charged-Kerr metric. Finally a summary and
discussion are given in section 5. In section 5 the comparison of
the scaling factors is also given.

\section{Symmetries and approximate symmetries of a Lagrangian}

The significance of variational symmetries is clear from the
celebrated Noether's theorem \cite{Nth}. According to this theorem
there is a procedure which relates the constants of the motion of a
given Lagrangian system to its symmetry transformations \cite{Ib,
KM}. Symmetry generators of a Lagrangian of a manifold form a Lie
algebra \cite{BKK}. Geometrically, KVs characterize the isometries of
a manifold \cite{HkEl}.

In general a manifold does not possess any exact symmetry but may do
so approximately. It is worth exploring the approximate symmetries
of a manifold, which form an approximate Lie algebra \cite{Gaz}.
Lie symmetries (and approximate Lie symmetries) of the system of the
geodesic equations for a spacetime yield conserved quantities but
there are also non-Noether symmetries that are not related to
conservation laws and therefore are of no interest for our purpose. To
calculate symmetries of a system of geodesic equation is tedious,
as it involves the second prolongation of the symmetry generator.
On the other hand the symmetries of a Lagrangian directly give us
the conserved quantities in which we are interested and here only the
first prolongation of the symmetry generator is required. Methods
for obtaining exact symmetries and first-order approximate symmetries
of a Lagrangian are available in the literature \cite{Ib, BKK, WF, TK}.
In this paper we extend the procedure of calculating the
approximate symmetries of a Lagrangian to the second order.

Noether symmetries, or symmetries of a Lagrangian, are defined as
follows. Consider a vector field defined on a real parameter fibre
bundle over the manifold \cite{Ib}
\begin{equation}
\mathbf {X}\mathbf{=}\xi (s,{x}^{\mu})\frac{\partial }{\partial
s}+\eta^{\nu}(s,{x}^{\mu})\frac{\partial}{\partial {x}^{\nu}},
\label{1}
\end{equation}
where $\mu,\nu=0,1,2,3$. The first prolongation of the above vector
field defined on the real parameter fibre bundle over the tangent
bundle to the manifold, is
\begin{equation}
\mathbf{X}^{[1]}=\mathbf{X}+(\eta^\nu_{,s}+\eta^\nu_{,\mu}\dot{x}^{\mu}
-\xi_{,s}\dot{x}^{\nu}-\xi_{,\mu}\dot{x}^{\mu} \dot{x}^{\nu})
\frac{\partial}{\partial \dot{x}^{\nu}}.\label{2}
\end{equation}
Generally one takes first-order Lagrangians as the corresponding
Euler-Lagrange equations are second-order ordinary differential
equations. In particular, we take $L(s,x^{\mu},\dot{x}^{\mu})$,
where $``\cdot"$ denotes differentiation with respect to the arc
length parameter $s$, which yields a set of second-order ordinary
differential equations (ODEs)
\begin{equation}
\ddot{x}^{\mu}=g(s,x^{\mu},\dot{x}^{\mu}). \label{3}
\end{equation}
Then {\bf X} is a Noether point symmetry of this Lagrangian if there
exists a gauge function, $A(s,x^{\mu})$, such that
\begin{equation}
\mathbf{X}^{[1]}L+(D_{s}\xi)L=D_{s}A,  \label{4}
\end{equation}
where
\begin{equation}
D_{s}=\frac{\partial }{\partial s}+\dot{x}^{\mu}\frac{\partial}
{\partial x^{\mu}}, \label{5}
\end{equation}
which is defined on the real parameter fibre bundle over the tangent
bundle to the manifold. For more general considerations and a
discussion of generalized symmetries see \cite{Ib, LMMR1}. The
significance of Noether symmetries is clear from the following
theorem \cite{Nth}.

{\bf Theorem 1}. If {\bf X} is a Noether point symmetry
corresponding to a Lagrangian $L(s,x^{\mu},\dot{x}^{\mu})$ of (3),
then
\begin{equation}
I={\xi}L+(\eta^{\mu}-\dot{x}^{\mu}{\xi})\frac{\partial L}{\partial
\dot{x}^{\mu}}-A, \label{6}
\end{equation}
is a first integral of (3) associated with {\bf X}. For the proof of
this theorem see for example \cite{Ovs}.

For a second-order (in $\epsilon$) perturbed system of ODEs
\begin{equation}
\mathbf{E}=\mathbf{E}_{0}+\epsilon\mathbf{E}_{1}+{\epsilon}^{2}
\mathbf{E}_{2}=O({\epsilon}^{3}), \label{7}
\end{equation}
{\it second-order approximate symmetries} of the first-order Lagrangian
\begin{equation}
L(s,x^{\mu},\dot{x}^{\mu},\epsilon)=L_{0}(s,x^{\mu},\dot{x}^{\mu})+
{\epsilon}L_{1}(s,x^{\mu},\dot{x}^{\mu})+{\epsilon}^{2}
L_{2}(s,x^{\mu},\dot{x}^{\mu})+O({\epsilon}^{3}),\label{8}
\end{equation}
are defined as follows. The functional ${\int}_{V}{L}ds$ is
invariant under the one-parameter group of transformations with
approximate Lie symmetry generator
\begin{equation}
\mathbf{X}=\mathbf{X}_{0}+\epsilon\mathbf{X}_{1}
+{\epsilon}^{2}\mathbf{X}_{2}+O({\epsilon}^{3}),\label{9}
\end{equation}
up to gauge
\begin{equation}
A={A}_{0}+\epsilon{A}_{1}+{\epsilon}^{2}{A}_{2},
\label{10}
\end{equation}
where
\begin{equation}
\mathbf{X}_{j}=\xi_{j}\frac{\partial}{\partial
s}+\eta^{\mu}_{j}\frac{\partial }{\partial
{x}^{\mu}},\;\;\;\;\;\;(j=0,1,2), \label{11}
\end{equation}
\begin{equation}
\mathbf{X}_{0}^{[1]}L_{0}+(D_{s}{\xi}_{0})L_{0}=D_{s}A_{0},
\label{12}
\end{equation}
\begin{equation}
\mathbf{X}_{1}^{[1]}L_{0}+\mathbf{X}_{0}^{[1]}L_{1}+(D_{s}{\xi}_{1})L_{0}
+(D_{s}{\xi}_{0})L_{1}=D_{s}A_{1}\label{13}
\end{equation}
and
\begin{equation}
\mathbf{X}_{2}^{[1]}L_{0}+\mathbf{X}_{1}^{[1]}L_{1}+\mathbf{X}_{0}^{[1]}L_{2}
+(D_{s}{\xi}_{2})L_{0}+(D_{s}{\xi}_{1})L_{1}+(D_{s}{\xi}_{0})L_{2}=D_{s}A_{2}.
\label{14} \end{equation}
For the first-order perturbed case (13) corresponding to a single equation,
see for example \cite{TK}.

Here $\mathbf{X}_{0}$ is the exact symmetry generator,
$\mathbf{X}_{1}$ is the first-order approximate part, $\mathbf{X}_{2}$
is the second-order approximate part of the approximate symmetry
generator, ${L}_{0}$ is the exact Lagrangian corresponding to the
exact equations $\mathbf{E}_0=0$, and
${L}_{0}+\epsilon{L}_{1}$ the first-order approximate Lagrangian
corresponding to the first-order perturbed equations
$\mathbf{E}_{0}+\epsilon\mathbf{E}_{1}=0$. The perturbed equations
(13) and (14) always have the approximate symmetry generators
$\epsilon \mathbf{X}_{0}$ which are known as ``trivial" approximate
symmetries and $\mathbf{X}$ given by (9) with $\mathbf{X}_{0}\neq 0$
is called a ``non-trivial" approximate symmetry.

\section{Symmetries and approximate symmetries of a Lagrangian for the
Kerr spacetime}

The Kerr spacetime is an axially symmetric, stationary solution of
the Einstein vacuum field equations. The line element for this
spacetime in Boyer-Lindqust coordinates is given by \cite{MTW}
\begin{equation}
ds^{2}=(1-\frac{2Gmr}{\rho^{2}{c}^2})c^2dt^{2}-(\frac{\rho^{2}}{\Delta})
dr^{2}-\rho^{2}d\theta^2-\Lambda\frac{\sin^{2}\theta}{\rho^{2}}d\phi^{2}
+(\frac{2Gmra\sin^{2}\theta}{\rho^{2}{c}^2})dtd\phi,\label{16}
\end{equation}
where
\begin{equation}
\rho^2=r^{2}+\frac{a^{2}}{c^2}\cos^{2}\theta,\quad
\Lambda=(r^{2}+\frac{a^{2}}{c^2})^{2}-\frac{a^{2}}{c^2}\Delta\sin^{2}\theta,
\quad \Delta=r^{2}+\frac{a^{2}}{c^2}-\frac{2Gmr}{c^2},\nonumber
\end{equation}
with $m$ the mass and $a$ the angular momentum per unit mass of the
gravitating source. This metric reduces to the Schwarzschild metric
when $a$ = 0. This spacetime has two KVs which give the energy and
azimuthal angular momentum conservation laws. Besides, there is a
non-trivial Killing tensor for this spacetime \cite{Car} which yields
the square of the total angular momentum \cite{AJM}.

We consider the Lagrangian for minimizing the arc-length (written
from the square of the arc length for convenience) which yields the
{\it geodesic equations} as the Euler-Lagrange equations,
\begin{equation}
L[x^{\mu},\dot{x}^{\mu}]=g_{\mu
\nu}(x^{\sigma})\frac{dx^{\mu}}{ds}\frac{dx^{\nu}}{ds}. \label{17}
\end{equation}
For the metric given by (15) it becomes
\begin{equation}
L=(1-\frac{2Gmr}{\rho^{2}{c}^2})c^2\dot{t}^{2}-\frac{\rho^{2}}{\Delta}\dot{r}^{2}
-\rho^{2}\dot{\theta}^{2}-\Lambda\frac{\sin^{2}\theta}{\rho^{2}}\dot{\phi}^{2}
+\frac{2Gmra\sin^{2}\theta}{\rho^{2}{c}^2}\dot{t}\dot{\phi}.
\label{18}
\end{equation}

Using (17) in (4) we obtain the 19 determining (partial
differential) equations for 6 unknown functions $\xi$, $\eta_{\mu}$
and $A$, where each of these is a function of 5 variables, i. e.
$s$, $t$, $r$, $\theta$ and $\phi$. Solving these equations we get
the isometries for the Kerr metric, the geodesic parameter
translation and the gauge function, i.e.
\begin{equation}
\mathbf{Y}_{0}=\frac{\partial}{\partial t},\quad
\mathbf{Y}_{3}=\frac{\partial}{\partial \phi},\quad
\mathbf{W}_{0}=\frac{\partial}{\partial s}\quad {\rm and}\quad
A=c\quad (\rm constant). \label{19}
\end{equation}
Thus, here we see that the isometries form a sub-algebra of the
symmetries of the Lagrangian. Use of (18) in (6) will provide the
first integrals of the geodesic equations for the Kerr metric.

For the approximate symmetries of a Lagrangian for the geodesic
equations in the Kerr spacetime we first consider the Kerr metric as
a first perturbation of the Schwarzschild metric by introducing the
spin angular momentum per unit mass $a/c^2$ as a small parameter
$\epsilon$. This first-order perturbed Lagrangian is given by
\begin{equation}
L=(1-\frac{2Gm}{r{c}^2})c^2\dot{t}^2-{(1-\frac{2Gm}{r{c}^2})}^{-1}\dot{r}^2
-r^2(\dot{\theta}^2+\sin^{2}{\theta}\dot{\phi}^2)+\epsilon\frac{2Gm}{r}
\sin{\theta}\dot{t}\dot{\phi}+O(\epsilon^2).\label{20}
\end{equation}
For $\epsilon=0$ we recover the Lagrangian of the unperturbed
Schwarzschild metric. The symmetry algebra of the Lagrangian is 5
dimensional, given by $so(3)\oplus\R\oplus{d}_{1}$, and it properly
contains the isometry algebra. The gauge function $A$ is just a
constant. From this information and (6) one can obtain the first
integrals of the geodesic equations for the Schwarzschild metric. Using
the $5$ exact symmetry generators in (13) we get the set of
determining equations whose solution gives us no non-trivial symmetry
but only exact symmetries are recovered as trivial first-order
approximate symmetries. Here we have recovered the conservation laws
of angular momentum as trivial first-order approximate conservation
laws which were lost in going from the Schwarzschild to the Kerr
spacetime.

Next we take the Kerr spacetime as a second perturbation of the
Minkowski spacetime. For this purpose we set
\begin{equation}
m=\epsilon \mu, a=\epsilon \alpha, \label{20}
\end{equation}
where $\mu={c^2}/{2G}$ and $\alpha=c\sqrt{k_{1}}$. For the Kerr
black hole (see, e.g. \cite{MTB}) we have $0<k_1\leq1/4$. Here the
second-order perturbed Lagrangian is given by
\begin{eqnarray}
L=\dot{t}^{2}-\dot{r}^{2}-r^{2}\dot{\theta}^{2}-r^{2}\sin^{2}\theta
\dot{\phi}^{2}-\frac{1}{r}\epsilon(\dot{t}^{2}+\dot{r}^{2})
-\epsilon^{2}[\frac{1}{r^{2}}(1-\frac {k_1^2}{4} \sin^2\theta)
\dot{r}^2 \nonumber \\
~~~~~~~~~~~+k_1^2 \cos^2\theta\dot{\theta}^2 +k_1^2
\sin^2\theta\dot{\phi}^2-\frac{\sqrt{k_1}}{r}\sin^2\theta
\dot{t}\dot{\phi}]+O(\epsilon^3). \label{21}
\end{eqnarray}
For the exact case, $\epsilon=0$, i.e. no mass or angular momentum
per unit mass, the Lagrangian (21) reduces to that of the Minkowski
spacetime. It has a 17 dimensional Lie algebra spanned by the
symmetry generators: 10 $\mathbf{Y}_i$'s, which are generators of
the Poincar\'{e} algebra $so(1,3)\oplus _{s}{\R}^{4}$,
\begin{align}
\mathbf{Y}_{0}&=\frac{\partial }{\partial t}, \quad
\mathbf{Y}_{1}=\cos \phi \frac{\partial }{\partial \theta }-\cot
\theta \sin \phi \frac{\partial }{\partial \phi },  \label{22}\\
\mathbf{Y}_{2}&=\sin \phi \frac{\partial }{\partial \theta }+\cot
\theta \cos \phi \frac{\partial }{\partial \phi },\quad
\mathbf{Y}_{3}=\frac{\partial }{\partial \phi },  \label{23}\\
\mathbf{Y}_{4}&=\sin \theta \cos \phi \frac{\partial }{\partial
r}+\frac{\cos \theta \cos \phi }{r}\frac{\partial }{\partial \theta
}-\frac{\csc \theta \sin \phi }{r}\frac{\partial }{\partial \phi },
\label{24}\\
\mathbf{Y}_{5}&=\sin \theta \sin \phi \frac{\partial }{\partial
r}+\frac{\cos \theta \sin \phi }{r}\frac{\partial }{\partial \theta
}+\frac{\csc \theta\cos \phi }{r}\frac{\partial }{\partial \phi },  \label{25} \\
\mathbf{Y}_{6}&=\cos \theta \frac{\partial }{\partial r}-\frac{\sin
\theta }{r}\frac{\partial }{\partial \theta },  \label{26}\\
\mathbf{Y}_{7}&=\frac{r\sin \theta \cos \phi }{c}\frac{\partial
}{\partial t}+ct(\sin \theta \cos \phi \frac{\partial }{\partial
r}+\frac{\cos \theta \cos \phi }{r}\frac{\partial }{\partial \theta
}-\frac{\csc \theta \sin \phi }{r}\frac{\partial }{\partial \phi }),
\label{27}\\
\mathbf{Y}_{8}&=\frac{r\sin \theta \sin \phi }{c}\frac{\partial
}{\partial t} +ct(\sin \theta \sin \phi \frac{\partial }{\partial
r}+\frac{\cos \theta \sin \phi }{r}\frac{\partial }{\partial \theta
}+\frac{\csc \theta \cos \phi
}{r}\frac{\partial }{\partial \phi }),  \label{28} \\
\mathbf{Y}_{9}&=\frac{r\cos \theta }{c}\frac{\partial }{\partial
t}+ct(\cos \theta \frac{\partial }{\partial r}-\frac{\sin \theta
}{r}\frac{\partial }{\partial \theta }),  \label{29}
\end{align}
and 7 other generators, whose significance is discussed below
\begin{align}
\mathbf{W}_{0}&=\frac{\partial}{\partial
s},\quad\mathbf{W}_{1}=s\frac{\partial}{\partial
s}+\frac{1}{2}(t\frac{\partial}{\partial
t}+r\frac{\partial}{\partial r}), \label{30}\\
\mathbf{Z}_{0}&=s\mathbf{Y}_{0},\quad
\mathbf{Z}_{1}=s\mathbf{Y}_{4},\quad
\mathbf{Z}_2=s\mathbf{Y}_{5},\quad \mathbf{Z}_3=s\mathbf{Y}_{6},
\label{31}\\
\mathbf{Z}_{4}&=\frac{1}{2}s(s\frac{\partial}{\partial
s}+t\frac{\partial}{\partial t}+r\frac{\partial}{\partial r}).
\end{align}
As before, the generator $\mathbf{W}_{0}$ gives translation in $s$
and always exists for a Lagrangian of the type (16) \cite{AQ},
$\mathbf{W}_{1}=[\mathbf{W}_{0},\mathbf{Z}_{4}]$ which is a scaling
symmetry in $s$,$t$,$r$ that can be used to get rid of the $s$
dependence in the generators given by (31) and (32). This is
reasonable as symmetries of a Lagrangian always form a sub-algebra
of the symmetries of the Euler-Lagrange (geodesic) equations
\cite{OLV} and the algebra of the Euler-Lagrange equations for
Minkowski spacetime is $sl(6,\R)$ which is 35 dimensional \cite{TQ}.
As mentioned above, using $\mathbf{W}_{1}$, we can write $s=t^2$ or
$s=r^2$ and
\begin{equation}
\mathbf{Z}_{4}=\frac{r^2}{4}[\frac{1}{t}(r^2+2t^2)\frac{\partial}
{\partial t}+3r\frac{\partial}{\partial r}].
\end{equation}
Now, every flat spacetime is conformally flat, i.e. for which all
components of the Weyl tensor are zero \cite{HkEl}. The Lie algebra
of the Conformal Killing Vectors (CKVs) for a conformally flat
spacetime is 15 dimensional \cite{hall}. Therefore for the Minkowski
spacetime we already know that there are 15 CKVs. The 5 symmetry
generators, i.e. $\mathbf{Z}_{i}$ for $i=0,...,4$ given by (31) and
(32), are proper CKVs with conformal factor $\psi=(c_0t^2+c_1)/2$.
Thus we see that not only the KVs but also the CKVs form a
sub-algebra of the symmetries of the Lagrangian for the Minkowski
spacetime. The extra 2 generators, $\mathbf{W}_0$,$\mathbf{W}_1$,
essentially provide the translation and appropriate scaling in the
geodetic parameter.

The gauge function is
\begin{equation}
A=\frac{1}{2}c_{0}(t^{2}-r^{2})+2tc_{3}+c_{4}-2r(c_{14}\sin\theta
\cos\phi+c_{15} \sin\theta\sin\phi+c_{16}\sin\theta), \label{32}
\end{equation}
where $c_{0},...,c_{16}$ are the arbitrary constants of integration
associated with the symmetry generators.

Retaining terms of first-order in $\epsilon$ and neglecting
$O(\epsilon^{2})$, the Lagrangian (21) becomes a first-order
perturbed Lagrangian for the Schwarzschild metric considered as a
first perturbation of the Minkowski metric. Using (13) and the
exact symmetry generators given by (22) - (32) we get a new set of
determining equations. In these equations only 12 of the 17 exact
symmetry generators appear. These 12 generators of the exact
symmetry have to be eliminated for consistency of these determining
equations, making them homogeneous. The resulting system is the same
as for the Minkowski spacetime, yielding 17 first-order approximate
symmetry generators given by (22) - (32). Thus for the Schwarzschild
metric as a first-order approximate case, we recover all the lost
conservation laws as trivial first-order approximate conservation
laws. Beside energy and angular momentum which always remain
conserved for the Schwarzschild metric (for both the exact and
perturbed cases) we see approximate conservation of linear momentum
and spin angular momentum. This was also observed for the first-order
approximate symmetries of the geodesic equations for the Schwarzschild
metric \cite{KMQ}.

Going from Minkowski to the Kerr spacetime we are left with only two
KVs which give conservation of energy and azimuthal angular
momentum. For the Lagrangian of the geodesic equations for exact
(unperturbed) Kerr metric there are only three symmetry generators
given by (18). We see that there is no non-trivial approximate
symmetry in the first-order approximation. To check whether we can
see non-trivial approximate symmetries from the definition of
second-order approximate symmetries of a Lagrangian, which will
hopefully give us a non-trivial conservation law, we take the Kerr
metric as a second-order perturbation of the Minkowski spacetime. In
the second approximation, that is when we retain terms quadratic in
$\epsilon$, we have the Lagrangian given by (21). From (14) we have
a new system of 19 determining equations. In these equations now 14
of the 17 exact (also first-order approximate) symmetry generators
appear. The two symmetry generators that arise here, which did not
occur in the set of determining equations for first-order
approximation, are $\mathbf{Y}_{1}$ and $\mathbf{Y}_{2}$ given in
(22) and (23). At first sight it seems that these two new symmetry
generators may yield some non-trivial second-order approximate
symmetries. But for the consistency of the determining equations all
the 14 constants have to be eliminated and the system again becomes
homogeneous. The resulting system is once more the same as for the
Minkowski spacetime, yielding 17 second-order approximate symmetry
generators given by (22) - (32). Thus there is no non-trivial
second-order approximate symmetry generator. In the second-order
approximation we recover all the lost conservation laws as trivial
second-order approximate conservation laws for the Kerr spacetime.
Hence we have recovered the Lorentz covariance using approximate
symmetries of the Lagrangian.

\section{Second-order approximate symmetries of the geodesic
equations for the charged-Kerr metric and rescaling of energy of a
test particle}

We have studied approximate symmetries of a Lagrangian for the Kerr
spacetime in which we recovered trivial first-order and second-order
approximate conservation laws. The rescaling of energy of test
particles was seen from the approximate symmetries of the geodesic
equations \cite{IMQ}. Therefore we now consider approximate
symmetries of the geodesic equations. In \cite{KMQ} the definition
of first-order approximate symmetries of DEs was used to calculate
first-order approximate symmetries of the geodesic equations for the
Schwarzschild metric. There, in the perturbed equations (given by
(40) here), instead of the perturbed system (given as subscript in
(40)) the exact system of geodesic equations was used and no energy
rescaling was forthcoming. The interesting result of energy
rescaling  of test particles for RN spacetime \cite{IMQ} was seen by
application of the definition of the second-order approximate
symmetries of DEs, wherein the perturbed system of geodesic
equations was used. It was further remarked that it should be
checked if the result of rescaling also holds for the Kerr metric.
Here we investigate this question.

In the RN metric the charge appears as a second-order perturbation
of the Minkowski metric \cite{IMQ}. The quadratic term in charge
appears in the scaling factor. Hence, here we investigate the
charged-Kerr metric to keep the charge up to the same second-order
and relate the scaling factor for this metric with that of the RN
spacetime.

In the charged-Kerr metric we have
\begin{equation}
g_{00}=1-\frac{G(2c^2mr-Q^2)}{{\rho}^2{c^4}},\quad
g_{03}=\frac{a}{{\rho}^2c^2}G(2mr-\frac{Q^2}{c^2})\sin^2{\theta}
\quad, \quad\Delta=\frac{a^2}{c^2}+r^2-\frac{G}{c^2}
(2mr-\frac{Q^2}{c^2}).
\end{equation}

Setting $Q=\epsilon \chi$, where $\chi=c^2 \sqrt{k/G}$ and
$\epsilon$ is defined by (20), we have the second-order approximate
geodesic equations for the Kerr metric
\begin{equation}
\ddot{t}+\epsilon\frac{1}{r^{2}}\dot{t}\dot{r}+\epsilon^{2}
[\frac{1}{r^{3}}(1-2k)\dot{t}\dot{r}
-\frac{2\sqrt{k_1}}{r^{2}}\sin^2{\theta}\dot{r}\dot{\phi}]+O(\epsilon)^{3}=0,
\label{33}
\end{equation}
\begin{equation}\begin{split}
\ddot{r}-r(\dot{\theta}^{2}+\sin^{2}\theta\dot{\phi}^{2})
+\epsilon[\frac{1}{2{r}^{2}}(\dot{t}^{2}
-\dot{r}^{2})+(\dot{\theta}^{2}+\sin^{2}\theta\dot{\phi}^{2})]
-\epsilon^{2}[\frac{1}{{2}r^{3}}(1+2k)\dot{t}^{2}
+\frac{\sqrt{k_1}}{r^{2}}\sin^{2}\theta\dot{t}\dot{\phi}&\\
-\frac{1}{r^3}(2(k_1\sin{\theta}+k)-1)\dot{r}^{2}+
\frac{k_1}{r^{2}}\sin^2\theta\dot{r}\dot{\theta}
+\frac{1}{r}(k_{1}\sin^2{\theta}+k)(\dot{\theta}^2+\sin^2{\theta}\dot{\phi}^2)]
+O(\epsilon)^{3}=0, \label{34}
\end{split}\end{equation}
\begin{equation}\begin{split}
\ddot{\theta}+\frac{2}{r}\dot{r}\dot{\theta}-\sin\theta\cos\theta\dot{\phi}^{2}
+\epsilon^{2}[\frac{\sqrt{k_1}}{r^{3}}\sin2\theta\dot{t}\dot{\phi}
-\frac{k_1}{{2}r^{4}}\sin2\theta\dot{r}^{2}
-\frac{2{k_1}}{r^{3}}\cos^2\theta\dot{r}\dot{\theta}\\
-\frac{k_1}{{2}r^{2}}\sin2{\theta}(\dot{\theta}^2
+\sin^2{\theta}\dot{\phi}^2)]+O(\epsilon)^{3}=0,\label{35}
\end{split}\end{equation}
\begin{equation}
\ddot{\phi}+\frac{2}{r}\dot{r}\dot{\phi}+2\cot\theta\dot{\theta}\dot{\phi}+
\epsilon^{2}[\frac{\sqrt{k_1}}{r^{4}}\dot{t}\dot{r}
+\frac{2\sqrt{k_1}}{r^{3}}\cot\theta\dot{t}\dot{\theta}
-\frac{2}{r^{3}}k_{1}\dot{r}\dot{\phi}] +O(\epsilon)^{3}=0.
\label{36}
\end{equation}
If $\epsilon=0$, these equations reduce to those of the
Minkowski metric. When we retain terms only up to order
$\epsilon$ and neglect higher orders, they reduce to
the first-order approximate geodesic equation of the Schwarzschild
metric. If we put $k_{1}=0$, they further reduce to those
of the RN-metric. We now apply the definition of
second-order approximate symmetries of a system of ODEs,
\begin{equation}
(\mathbf{X}_{0}+\epsilon\mathbf{X}_{1}+\epsilon^{2}\mathbf{X}_{2})\left(\mathbf{E}_{0}
+\epsilon\mathbf{E}_{1}+\epsilon^{2}\mathbf{E}_{2})\right\vert
_{\mathbf{E}_
{0}+\epsilon\mathbf{E}_{1}+\epsilon^{2}\mathbf{E}_{2}=O(\epsilon^{3})}=O(\epsilon^{3}),
\label{37}
\end{equation}
(see \cite{IMQ} and references given there in) to (36) - (39), where
$\mathbf{X}_{0}$ is the exact symmetry generator, $\mathbf{X}_{1}$,
$\mathbf{X}_{2}$ are the first-order and second-order approximate
parts of the approximate symmetry generator respectively,
$\mathbf{E}_{0}$ is the exact part, $\mathbf{E}_{1}$ is the
first-order perturbed part and $\mathbf{E}_{2}$ is the second order
perturbed part of the system of ODEs respectively. The exact
symmetry algebra includes the generators of the dilation algebra,
${\partial}/{\partial s}$, $s{\partial}/{\partial s}$ corresponding
to
\begin{equation}
\xi (s)=c_{0}s+c_{1}.   \label{38}
\end{equation}

In the determining equations for the first-order approximate
symmetries \cite{KMQ} the terms involving $\xi_{s}=c_{0}$ cancel
out. Taking the RN metric as a second perturbation of the Minkowski
metric \cite{IMQ}, it was seen that the terms involving $\xi_{s}$ do
not automatically disappear but collect a scaling factor of
$(1-Q^{2}/2Gm^{2})$ in order to cancel out. In the case of the
charged-Kerr spacetime, as a second perturbation of the Minkowski
spacetime, the terms involving $\xi_{s}$ in the set of determining
equations also do not disappear automatically but collect a scaling
factor
\begin{equation}
(1/r^3)(1-2k)\dot{t}-(2/r^2)(\sqrt{k_1}\sin^2{\theta})\dot{\phi},
\end{equation}
so as to cancel out, where $k=Q^2/4Gm^2$ and $k_{1}=a^2c^2/4G^2m^2$.
From (1) one can see that $\xi$ is the coefficient of
${\partial}/{\partial s}$ in the point transformations. This scaling
factor involves the derivatives of the coordinates $t$ and $\phi$,
which can be replaced by the first integrals of the geodesic
equations and involve constants that are the mass and the spin. As
such, we put them in as $m$ and $a$. Thus we get (taking $G=1$, $c=1$)
\begin{equation}
M_{c-K}=m-\frac{Q^2}{2m}+\frac{ma}{2r}.
\end{equation}
For $a=0$, (43) reduces to $m$-times of the expression for the RN
spacetime \cite{IMQ}.

Komar, using his definition of approximate symmetry \cite{AK}, wrote
down an integral for the mass in a spacetime \cite{AK1}
\begin{equation}
M=\frac{1}{8\pi}{\int}_{s^2}{*d\tilde{\xi}},
\end{equation}
where $\tilde{\xi}$ is the time-like Killing 1-form for the exact
symmetry, ${*d\tilde{\xi}}$ the dual of the 2-form ${d\tilde{\xi}}$
and $s^2$ is the 2-surface \cite{CF, DC, LS}. Using the Komar
integral (44) Cohen and de Felice considered $\xi$ as the stationary
Killing 1-form over a charged-Kerr background metric \cite{CF}. They
obtained a formula for the effective mass (and hence energy) for the
charged-Kerr spacetime
\begin{equation}
M_{c-K}= m-\frac{Q^2}{r}-\frac{Q^2(r^2+a^2)}{ar^2}\tan^{-1}(\frac{a}{r}).
\end{equation}
In the above expression (45) $a$ does not appear explicitly and only
appears in a product with $Q$. When $Q\longrightarrow 0$ in the
above expression (45) the effects of rotation also disappear. This
does not seem reasonable. In the limit of $a\longrightarrow  0$
expression (45) reduces to that of the RN spacetime given in
\cite{CG, VW}.

Chellathurai and Dadhich modified the Komar integral and obtained an
expression for the effective mass of the charged-Kerr black hole \cite{DC}
\begin{equation}
M_{c-K}=m-\frac{Q^2}{r}-\frac{(12m^2+Q^2)a^2}{3r^3}+\frac{14ma^2 Q^2}{3r^4}+....
\end{equation}
This expression (46) reduces to that of the RN spacetime in the limit
$a\longrightarrow 0$ and in the limit $Q\longrightarrow 0$ reduces to
that for the Kerr spacetime \cite{RCD}. However, it is not clear that
this modification satisfactorily adjusts for the approximate symmetry of
Komar.

Qadir and Quamar \cite{AJ} obtained an expression for the
$\psi N$-potential of the charged-Kerr spacetime,
\begin{equation}
\varphi =-\frac{mr-Q^2/2}{(r^2+a^2\cos^2\theta)}.
\end{equation}
In the limit $a\longrightarrow 0$ (47) reduces to that for the RN
spacetime \cite{Q1,Q2}. This yields the approximate modification of the
mass to be
\begin{equation}
M_{c-K}=m-\frac{Q^2}{2r}-\frac{ma^2\cos^2\theta}{r^2}+\frac{a^2 Q^2\cos^2\theta}{2r^3}+....
\end{equation}

The significance and comparison of our expression with (45), (46) and
(48) will be discussed further in the next section.

\section{Summary and Discussion}

In this paper we have discussed exact and approximate symmetries of
a Lagrangian for the geodesic equations in the Kerr spacetime.
Minkowski spacetime is maximally symmetric having 10 KVs. Going from
Minkowski to the Kerr spacetime we are left only with two KVs which
correspond to energy and azimuthal angular momentum conservation.
The unperturbed Lagrangian for the geodesic equations in the Kerr
spacetime has an additional symmetry ${\partial}/{\partial s}$ and
the unperturbed Lagrangian for the Schwarzschild metric has a 5
dimensional algebra which contains the four KVs of this metric and
${\partial}/{\partial s}$. Taking the Kerr spacetime as a first
perturbation of the Schwarzschild metric with spin as a small
parameter we recovered the conservation laws as trivial first-order
approximate conservation laws which were lost in going from the
Schwarzschild spacetime to the Kerr spacetime.

Retaining terms of $O(\epsilon^{2})$ in the Kerr spacetime we have a
second-order perturbed Lagrangian given by (21). This Lagrangian
reduces to that of Minkowski spacetime if $\epsilon=0$ and if we
retain terms of first-order in $\epsilon$ and neglecting
$O(\epsilon^{2})$, we get a Lagrangian for the perturbed
Schwarzschild metric which is a first perturbation of the Minkowski
metric. For the exact case (Minkowski spacetime) symmetries of the
Lagrangian form a 17 dimensional Lie algebra, which also holds in
Cartesian coordinates and thus there is no coordinate dependence.
[It may be mentioned here that the symmetries of the Minkowski
metric Lagrangian were first discussed in \cite{BKK}, where the
metric taken was $ds^{2}=\cosh(x/a)dt^{2}-dx^{2}-dy^{2}-dz^{2},$
which is not Minkowski, as it has ${R}^{0}_{101}\neq 0$. The
calculation was left incomplete, giving an impression that the
algebra is infinite dimensional, and it was shown that the isometry
algebra is a sub-algebra of the symmetries of the Lagrangian. We
pointed these errors out to the authors. This problem was revisited
in \cite{BKK1} with the correct metric, but the symmetry algebra of
the Lagrangian was given as 12 dimensional and the gauge function as
zero, which was again erroneous.]

For the first-order approximate case (perturbed Schwarzschild) there
is no non-trivial first-order approximate symmetry of the
Lagrangian. However all the exact 17 symmetry generators are
recovered as first-order approximate symmetry generators. In the
second-order approximate case, i.e. when we retain terms quadratic
in $\epsilon$, which is the second perturbation of the Minkowski
metric, we again have no non-trivial second-order approximate
symmetry of the Lagrangian and only 17 symmetry generators of the
exact case are recovered as second-order approximate symmetry
generators. Thus we see that in going from Minkowski to
Schwarzschild and Kerr metrics the conservation laws which were lost
are now recovered as approximate conservation laws. It was shown
\cite{AQ} that a Lagrangian possesses at least one additional
symmetry generator, ${\partial}/{\partial s}$, apart from the
isometry algebra. This is verified for the Schwarzschild and Kerr
spacetimes. As in the case of the Minkowski metric the CKVs form a
sub-algebra of the symmetries of the Lagrangian which include
${\partial}/{\partial s}$. We conjecture that {\it the CKVs form a
sub-algebra of the symmetries of the Lagrangian that minimize the
arc length, for any spacetime}.

For both the Schwarzschild and Kerr spacetimes the unperturbed
Lagrangian has only the one additional symmetry
${\partial}/{\partial}s$. For both the metrics the gauge function
$A$ is a constant. It remains an open question, whether this is true
in general for all 4 dimensional curved spacetimes. In Minkowski
spacetime there are 7 additional symmetries and the gauge function
$A$ is a function of 4 variables $t,r,\theta$ and $\phi$ given by
(34). In these additional 7 symmetry generators of the Minkowski
metric Lagrangian, which are also recovered as first-order and
second-order approximate symmetries generators for the Schwarzschild
and Kerr metrics respectively, $\mathbf{W}_{0}$ is the translation
in the geodetic parameter $s$ and $\mathbf{W}_{1}$ is used to
replace $s$ by $t^2$ in $\mathbf{Z}_{i}$, ($i=0,...,4$) to obtain
the CKVs. In the exact (unperturbed) case, the symmetries of a
Lagrangian form a sub-algebra of symmetries of the Euler-Lagrange
equations \cite{OLV}. Here we conjecture that {\it approximate
symmetries of a perturbed Lagrangian also form a sub-algebra of the
approximate symmetries of the perturbed Euler-Lagrange equations}.

We also looked at the second-order approximate symmetries of the
geodesic equations for the charged-Kerr spacetime to find a
rescaling factor. Since the rescaling comes in the derivative
relative to proper time, it was argued \cite{IMQ} that it gives a
rescaling of the energy in this spacetime. In the RN spacetime
\cite{IMQ}, the rescaling was independent of $r$ while for the
charged-Kerr metric the rescaling factor given by (43) consists of
two parts - one is due to charge and the other is due to spin of the
gravitating source which depends on $r$. The charge comes in {\it
quadratically} compared to unity in one term. The spin comes in {\it
linearly}. It does not come with a constant term to compare.
However, taken as a whole, we see that the spin has an effectively
{\it lower order} effect.

In all three expressions (45), (46) and (48), the charge and spin
appear at the same order (quadratically). The last one comes with a
$\theta$-dependent part, which arises from the $\theta$-dependence
of the ``force" experienced by a body in the Fermi-Walker frame
\cite{QZ}. This $\theta$-dependence does not seem reasonable for the
defining the energy in the Kerr spacetime. As mentioned earlier,
(45) seems unreasonable as the rotational effect depends on the
presence of a charge! In (43) in the absence of charge, the effect
is to {\it enhance} the mass. This seems reasonable as the
frame-dragging effect also appears to lead to an enhanced mass -
``friction" of the rotating mass with the background spacetime, as
it were. Recall that one can extract rotational energy from a
rotating black hole and hence the rotation should {\it add} into the
mass. As would be expected, this effect decreases with $r$. The
other three expressions give a {\it reduction} of the rotating mass.
Also notice that (43) gives a change in the mass due to charge that
is position independent. That this should be so is not so clear to
us. However, nor is it clear to us that it {\it should} be position
dependent. The force experience by a particle in the field of a
charged gravitational source would be position dependent, but this
does not say that the mass should be modified by a position
dependent expression. It might be that in (43) the modification is
due to the electromagnetic self-energy to the gravitational
self-energy. As such, we conclude that the other three expressions
have definite drawbacks to be considered reliable and that (43)
seems to be free of those problems.

It would be of interest to analyse the Kerr-AdS and other solutions
using approximate Noether symmetries. One could use references
\cite{LMM} and \cite{LMM1} and those cited therein for the purpose.
In particular, there is no good definition of energy for spacetimes
containing gravitational waves, because of the lack of a timelike
KV. There is a proposal for a definition using superpotentials
\cite{LMMM, MM}, whose relationship to the definition using
approximate symmetries would be worth exploring. It is of interest
to apply this method of approximate symmetries of a Lagrangian to
gravitational waves in the hope of finding an approximate timelike
KV which will give energy conservation up to a certain
approximation. This matter will be discussed in detail elsewhere. A
preliminary discussion is given in \cite{IMQ}.

\section*{Acknowledgments} IH would like to thank Higher Education
Commission of Pakistan (HEC) for their full financial support and
DECMA of the University of Witwatersrand, Johannesburg, where the
writing-up was completed.

%\section*{ References }


\begin{thebibliography}{99}
\bibitem{MTW} Misner CW, Thorne KS and Wheeler JA, \textit{Gravitation},
 W.H. Freeman and Company, San Francisco, 1973.

\bibitem{LMMM1} Fatibene L, Ferraris M, Francaviglia M and Raiteri M, {\it J. Math.
Phys.}, {\bf 42} (2001) 1173 - 1195.

\bibitem{LMM2} Fatibene L, Ferraris M and Francaviglia M, {\it Int. J. Geom. Methods
Mod. Phys.}, {\bf 2} (2005) 373 - 392.

\bibitem{AK} Komar A, \textit{Phys. Rev}., \textbf{127} (1962) 1411 - 1418;
\textbf{129} (1963) 1873 - 1876.

\bibitem{JM} Matzner R, \textit{J. Math. Phys}., \textbf{9} (1968) 1657 - 1668; 1063 -
1067.

\bibitem{Ib} Ibragimov NH, \textit{Elementary Lie Group Analysis and Ordinary
Differential Equations}, Wiely, Chichester, 1999.

\bibitem{KMQ} Kara AH, Mahomed FM and Qadir A, \textit{Nonlinear Dyn}., \textbf{51} (2008) 183 -
188.

\bibitem{IMQ} Hussain I, Mahomed FM and Qadir A, \textit{SIGMA} \textbf{3}, (2007) 115, 9 pages,
arXiv.0712.1089.

\bibitem{AK1} Komar A, \textit{Phys. Rev}., \textbf{113} (1959) 934 -
936.

\bibitem{ESEFEs} Kramer D, Stephani H, MacCullum MAH and Herlt E, \textit{Exact
Solutions of Einstein Field Equations}, Cambridge University Press,
Cambridge, 1980.

\bibitem{AQs} Qadir A, ``Spacetime symmetries and their significance", in {\it Applications of
Symmetry Methods}, pp. 45 - 71, eds. Qadir A, and Saifullah K, National Centre for Physics 2006.

\bibitem{Lie} Lie S, Sophus Lie's 1880 \textit{Transformation Group} Paper, Math Sci.
Press, Brookline MA 1975. Translated by Michael Ackerman. Comments
by Robert Hermann.

\bibitem{CF} Choen JM and de Felice F, \textit{J. Math. Phys}., \textbf{25}
(1984) 992 - 994.

\bibitem{DC} Chellathurai V and Dadhich N, \textit{Class. Quantum Grav}.
\textbf{7} (1990) 361 - 370.

\bibitem{AJ} Qadir A and Quamar J, \textit{Europhys. Lett}., \textbf{2} (1986)
423 - 425; \\
Qadir A, \textit{Europhys. Lett}., \textbf{2} (1986) 427 - 430.

\bibitem{Nth} Noether E, ``Invariant variations problems", \textit{Nachr. Konig.
Gissell. Wissen., Gottingen, Math.-Phys.Kl}. \textbf{2} (1918) 235.
(English translation in transport theory and Statistical Physics
\textbf{1} (1971)) 186.

\bibitem{KM} Kara AH and Mahomed FM, \textit{J. Nonlinear Math. Phys}., \textbf{9}
(2002) 60 - 72.

\bibitem{BKK} Bokhari AH, Kara AH, Kashif AR and Zaman FD,
\textit{Int. J. Theo. Phys}., \textbf{45} (2006) 1029 - 1039.

\bibitem{HkEl} Hawking SW and Ellis GFR, \textit{The Large Scale Structure
of Spacetime}, Cambridge University Press, Cambridge, 1973.

\bibitem{Gaz} Gazizov RK, \textit{J. Nonlinear Math. Phys}., \textbf{3} (1996) 96 -
101.

\bibitem{WF} Wafo Soh C and Mahomed FM, \textit{Class. Quantum Grav}. \textbf{16} (1999)
3553 - 3566.

\bibitem{TK} Feroze T and Kara AH, \textit{Inernational. J. Non-linear
Mechanics} \textbf{37} (2002) 275 - 280.

\bibitem{LMMR1} Fatibene L, Ferraris M, Francaviglia M and McLenaghan RG, {\it Int. J. Geom. Methods
Mod. Phys.}, {\bf 43} (2002) 3147 - 3161.

\bibitem{Ovs} Ovsiannikov LV, \textit{Group Analysis of Differential Equations}, New
York: Academic Press, 1980.

\bibitem{Car} Carter B, \textit{Commun. Math. Phys}. \textbf{10}
(1968) 280 - 310.

\bibitem{AJM} Qadir A, Quamar J and Rafique M, \textit{Phys. Lett.
A}, \textbf{109} (1985) 90 - 92.

\bibitem{MTB} Chandrasekhar S, \textit{The Mathematical Theory of Black Holes},
Clerendon Press Oxford, Oxford University Press, 1983.

\bibitem{AQ} Qadir A, \textit{SIGMA}. \textbf{3} (2007) 103, 7 pages, arXiv.0711.0814.

\bibitem{OLV} Olver PJ, \textit{Applications of Lie Groups to Differential
Equations}, Springer-Verlag, New York, 1993.

\bibitem{TQ} Feroze T, Mahomed FM and Qadir A, \textit{Nonlinear Dyn.}, \textbf{45} (2006) 65 -
74.

\bibitem{hall} Hall GS, \textit{Symmetries and Curvature Structure in
General Relativity}, World Scientific, Singapore, 2004.

\bibitem{LS} Szabados LB, \textit{Living Rev. Relativity}, \textbf{7} (2004) 1 -
140.

\bibitem{CG} Cohen JM and Gautreau R, \textit{Phys. Rev. D}., \textbf{19}
(1979) 2273 - 2279.

\bibitem{VW} de La Cruz V and Israel W, \textit{Nuovo Cimento}, \textit{51}
(1967) 744 - 759.

\bibitem{RCD} Kulkarni R, Chellathurai V and  Dadhich N, \textit{Class. Quantum Grav},
\textbf{5} (1988) 1443 - 1445.

\bibitem{Q1} Qadir A, ``General relativity in terms of forces",
in \textit{Proceedings of Third Regional Conference on Mathematical
Physics}, pp. 481-490, eds. Hussain F, and Qadir A, World scientific
1990.

\bibitem{Q2} Qadir A, ``The gravitational force in general relativity", in
\textit{M. A. B. Beg Memorial Volume}, pp. 159-178, eds. Ali A, and Hoodbhoy P,
World Scientific 1991.

\bibitem{BKK1} Bokhari AH, Kara AH, Kashif AR and Zaman FD,
\textit{Int. J. Theo. Phys}., \textbf{46} (2007) 2795 - 2800.

\bibitem{QZ} Qadir A and Zafarullah I, {\it Nuovo Cimento} {\bf B 111} (1996) 79 -
84.

\bibitem{LMM} Fatibene L, Ferraris M and Francaviglia M, {\it J. Geom. Methods Mod.
Phys.}, {\bf 3} (2006) 1341 - 1347.

\bibitem{LMM1} Fatibene L, Ferraris M and Francaviglia M, {\it J. Geom. Methods Mod.
Phys.}, {\bf 5} (2008) 1065 - 1068.

\bibitem{LMMM} Fatibene L, Ferraris M and Francaviglia M and Raiteri M, {\it Ann.
Physics}, {\bf 275} (1999) 27 - 53.

\bibitem{MM} Ferraris M and Francaviglia M,  {\it General Relativirt and Gravitatuinal Physics},
pp. 183 -196, eds. Teaneck NJ, World Scientific, 2004.

\end{thebibliography}
\end{document}